\documentclass{article}

\usepackage[round]{natbib}
\usepackage{amssymb}
\usepackage{mathrsfs}
\usepackage{bbm}

\usepackage{amsthm}
\usepackage{amsmath}
\usepackage{natbib}
\usepackage[colorlinks,citecolor=blue,urlcolor=blue,filecolor=blue,backref=page]{hyperref}
\usepackage{graphicx}

\usepackage[ruled]{algorithm2e}

\usepackage{booktabs}       
\usepackage{multirow, makecell}

\usepackage{fancyhdr}
\DeclareMathOperator*{\minimize}{minimize}
\DeclareMathOperator*{\argmin}{arg\,min}

\begin{document}

	\title{Convex Covariate Clustering for Classification}
	
	\author{Daniel Andrade (andrade@nec.com) \\
		\footnotesize{Central Research Laboratories, NEC} \\ \\
		%
		Kenji Fukumizu \\
		\footnotesize{The Institute of Statistical Mathematics} \\ \\
		%
		Yuzuru Okajima \\
		\footnotesize{Central Research Laboratories, NEC}}
	
	\date{}
	
	\maketitle
	\thispagestyle{fancy}
	
\begin{abstract}
Clustering, like covariate selection for classification, is an important step to compress and interpret the data.
However, clustering of covariates is often performed independently of the classification step,  
which can lead to undesirable clustering results that harm interpretability and compression rate.
Therefore, we propose a method that can cluster covariates while taking into account class label information of samples.
We formulate the problem as a convex optimization problem which uses both, a-priori similarity information between covariates, and information from class-labeled samples.
Like ordinary convex clustering \citep{chi2015splitting}, the proposed method offers a unique global minima making it insensitive to initialization.
In order to solve the convex problem, we propose a specialized alternating direction method of multipliers (ADMM), which scales up to several thousands of variables.
Furthermore, in order to circumvent computationally expensive cross-validation, we propose a model selection criterion based on approximating the marginal likelihood.
Experiments on synthetic and real data confirm the usefulness of the proposed clustering method and the selection criterion. 
\end{abstract}

\section{Introduction}

Interpretability is paramount to communicate classification and regression results to the domain expert. 
Especially in high-dimensional problems, sparsity of the solution is often assumed to increase interpretability. 
As a consequence, various previous work in statistics focuses on covariate selection, with the $\ell_1$-penalty being a particular popular choice \citep{hastie2015statistical}.

However, even after successful covariate selection, the remaining solution might still contain several hundreds or thousands of covariates.
We therefore suggest to further cluster the covariates.
In particular, we propose a clustering method that uses a-priori similarity information between covariates while respecting the response variable of the classification problem.
The resulting covariate clusters can help to identify meaningful groups of covariates and this way, increase the interpretability of the solution.

As a motivating example, consider the situation of document classification, where we are interested in classifying into positive and negative movie reviews. Assuming, a logistic regression model,
words with similar regression coefficients might convey similar sentiment, e.g. "wonderful" and "great", and therefore can be considered as one cluster.  On the other hand, we might also have prior word similarity information from word embeddings \citep{mikolov2013distributed} that can help us to identify meaningful clusters.


In order to incorporate, both types of information, class labels and a-priori covariate similarities, we propose to formulate the clustering problem as a joint classification and covariate clustering problem. In particular, we use a logistic regression loss with a pair-wise group lasso penalty for each pair of covariate regression coefficient vector.
Our formulation leads to a convex optimization problem, like convex clustering \citep{chi2015splitting,hocking2011clusterpath}.
As a consequence, we have the desirable properties of a unique global minima, and the ease to plot a clustering hierarchy, instead of just a single clustering. 

Our proposed method is conceptually related to joint convex covariate clustering and linear regression as proposed in \citep{she2010sparse}.
However, the change from linear regression to logistic regression is computationally challenging.
The main reason is that the objective function is not decomposable for each covariate anymore. 
%
%
Therefore, we propose a specialized alternating direction method of multipliers (ADMM) with an efficient gradient descent step for the non-decomposable primal update. 
Our solution allows us to scale the covariate clustering to problems with several 1000 covariates.

Since we often want to decide on \emph{one} clustering result, we also propose a model selection criterion using an approximation to the marginal likelihood. 
The motivation for this criterion is similar to the Bayesian information criterion (BIC) \citep{schwarz1978estimating}, but prevents the issue of a singular likelihood function. 
By using the marginal likelihood, we can circumvent the need for hyper-parameter selection with cross-validation, which is computationally infeasible when the number of covariates is large.

The outline of this article is as follows. In the next section, we introduce our proposed objective function, and describe an efficient ADMM algorithm for solving the convex optimization problem.
In Section \ref{sec:approxBMS}, we describe our model selection criterion for selecting a plausible clustering result out of all clustering results that were found with the proposed method.
In our experiments, in Sections \ref{sec:experimentsSyntheticData} and \ref{sec:experimentsRealData}, we compare the proposed method to a two-step approach, first step: covariate clustering with $k$-means clustering or convex clustering \citep{chi2015splitting}; second step: classification.
%

Finally, in Section \ref{sec:conclusions}, we summarize our conclusions.


A note on our notation: we denote a matrix by capital letter, e.g. $B \in \mathbb{R}^{c \times d}$, and a column vector by bold font e.g. $\mathbf{x} \in \mathbb{R}^{d}$. Furthermore, the i-th row of $B$ is denoted by $B_{i, \cdot}$ and is a row vector. The j-th column of $B$ is denoted by $B_{\cdot, j}$ or simply $\mathbf{b}_j$, and is a column vector. For a vector $\mathbf{x}$, we denote by $||\mathbf{x}||_2$ the $\ell_2$ norm, and for a matrix $B$, $||B||_F$ denotes the
Frobenius norm.

\section{Proposed Method} \label{sec:convexFormulation}

Let $B \in \mathbb{R}^{c \times d}$, where $c$ is the number of classes, and $d$ is the number of covariates.
$B_{l, \cdot}$ is the weight vector for class $l$. Furthermore, $\boldsymbol{\beta}_0 \in \mathbb{R}^{c}$ contains the intercepts.
We assume the multi-class logistic regression classifier defined by
\begin{align*} 
f(y | \mathbf{x} , B, \boldsymbol{\beta}_0) = \frac{\text{exp}(B_{y, \cdot} \mathbf{x} + \boldsymbol{\beta}_0(y))} {\sum_{y'} \text{exp}(B_{y', \cdot} \mathbf{x}  + \boldsymbol{\beta}_0(y'))} \, .
\end{align*}
\normalsize
We propose the following formulation for jointly classifying samples $\boldsymbol{x}_s$ and clustering the covariates:
\begin{align} \label{eq:convexSimpleFormulation}
\minimize_{B,\boldsymbol{\beta}_0}  \; \;  & - \sum_{s = 1}^n \log f(y_s | \mathbf{x}_s , B, \boldsymbol{\beta}_0) 
+ \lambda ||B||_F^2 \nonumber \\
& \quad + \nu \sum_{i_1 < i_2} S_{i_1, i_2} ||B_{\cdot, i_1} - B_{\cdot, i_2}||_2 \, ,
\end{align}
\normalsize
where $S_{i_1,i_2} \geq 0$ defines a similarity measure between covariate $i_1$ and $i_2$ and is assumed to be given a-priori.
The last term is a group lasso penalty on the class weights for any pair of two covariates $i_1$ and $i_2$. The penalty is large for similar covariates, and therefore encourages that $B_{\cdot, i_1} - B_{\cdot, i_2}$ is $\mathbf{0}$, that means that $B_{\cdot, i_1}$ and $B_{\cdot, i_2}$ are equal.
The clustering of the covariates can be found by grouping two covariates $i_1$ and $i_2$ together if  
$B_{\cdot, i_1}$ and $B_{\cdot, i_2}$ are equal. 


The advantage of this formulation is that, as long as $\lambda > 0$, the model is identifiable, and the optimization problem \eqref{eq:convexSimpleFormulation} is \emph{strongly} convex, and we are therefore guaranteed to find the unique global minima.\footnote{On the other hand, if $\lambda = 0$, and, for example, the optimal solution for $B$ has all entries equal to some constant, then the problem in Equation \eqref{eq:convexSimpleFormulation} is not strongly convex, but only convex. In practice, a value $\lambda > 0$ can potentially help to speed up convergence, and improve numerical stability. For all our experiments we set $\lambda$ to $0.1$.}

Note that this penalty shares some similarity to convex clustering as in \citep{chi2015splitting,hocking2011clusterpath}.
However, one major difference is that we do not introduce latent vectors for each data point, and our method can jointly learn the classifier and the clustering.

We remark that it is straight forward to additionally add a sparsity penalty on the columns of $B$ to jointly select and cluster all covariates. We omit in the following such extensions and focus on the computationally difficult part, the clustering.
Our implementation\footnote{Released here \url{https://github.com/andrade-stats/convexCovariateClusteringClassification}.} allows to perform joint or sequential covariate selection and clustering.

Finally, we remark that the similarity matrix can be represented as an undirected weighted graph with an edge between covariates $i_1$ and $i_2$ iff $S_{i_1,i_2} > 0$. 
In practice, $S$ might be manually crafted from a domain expert (e.g. given by some ontology), or learned from data (as we do in Sections \ref{sec:experimentsSyntheticData} and \ref{sec:experimentsRealData}).

\subsection{Optimization using ADMM}

Here, we focus on the optimization problem from Equation \eqref{eq:convexSimpleFormulation}, which we can rewrite as
\begin{align*}
\minimize_{B,\boldsymbol{\beta}_0} \; \;  & - \sum_{s = 1}^n \log \tilde{f}(y_s | \mathbf{x}_s , B, \boldsymbol{\beta}_0) \\
&\quad  + \nu \frac{1}{2} \sum_{i = 1}^{d} \sum_{j = 1}^{d_i} S_{i, e_i(j)} ||B_{\cdot, i} - B_{\cdot, e_i(j)}||_2  \, , 
\end{align*}
\normalsize
where $d_i$ denotes the number of adjacent covariates of $i$, and 
\begin{align*}
\tilde{f}(y_s | \mathbf{x}_s , B, \boldsymbol{\beta}_0) := f(y_s | \mathbf{x}_s , B, \boldsymbol{\beta}_0) \cdot \text{exp}(- \frac{\lambda}{n} ||B||_F^2) \, .
\end{align*}
\normalsize
Assuming that the adjacent covariates of $i$ are ordered from 1 to $d_i$, the function $e_i(j)$ returns the global covariate id of the $j$-th adjacent covariate of $i$.
We can then formulate the problem as
\begin{align*} 
\minimize_{B, Z, \boldsymbol{\beta}_0} \; \;  & - \sum_{s = 1}^n \log \tilde{f}(y_s | \mathbf{x}_s , B, \boldsymbol{\beta}_0) \\
& \quad + \nu \frac{1}{2} \sum_{i = 1}^{d} \sum_{j = 1}^{d_i} S_{i, e_i(j)} ||\mathbf{z}_{i \rightarrow e_i(j)} - \mathbf{z}_{e_i(j) \rightarrow i}||_2 \\
& \text{subject to} \; \;  \\ 
& \forall i \in \{1,\ldots d\}, j \in \{1, \ldots, d_i\}: \mathbf{z}_{i \rightarrow e_i(j)} - \mathbf{b}_i = \mathbf{0} \, , 
\end{align*}
\normalsize
where we denote $\mathbf{b}_j := B_{\cdot,j} \in  \mathbb{R}^{c}$.
Therefore, $\mathbf{z}_{i \rightarrow a}$ can be read as ``a copy of $\mathbf{b}_i$ for the comparison with $\mathbf{b}_a$", where $a$ is an adjacent node of $i$.

Using ADMM this can be optimized with the following sequence: 
\begin{align*} 
& B^{k+1} := \argmin_{B, \boldsymbol{\beta}_0} \; \; g(B, \boldsymbol{\beta}_0) \, , \\
%
& Z^{k+1}_{\cdot \rightarrow \cdot} := \argmin_{Z_{\cdot \rightarrow \cdot}} \; \; 
\nu \frac{1}{2} \sum_{i = 1}^{d} \sum_{j = 1}^{d_i} S_{i, e_i(j)} ||\mathbf{z}_{i \rightarrow e_i(j)} - \mathbf{z}_{e_i(j) \rightarrow i}||_2 \\
&\quad \quad \quad \quad \quad \quad \; \; + \frac{\rho}{2} \sum_{i = 1}^{d} \sum_{j = 1}^{d_i} ||\mathbf{z}_{i \rightarrow e_i(j)} - \mathbf{b}_{i}^{k+1} +  \mathbf{u}_{i \rightarrow e_i(j)}^k ||_2^2 \, , \\
%
%
& \forall \{i, j\} : i \in \{1,\ldots d\}, j \in \{1, \ldots, d_i\}: \\
& \mathbf{u}_{i \rightarrow j}^{k+1} := \mathbf{u}_{i \rightarrow j}^{k} + \mathbf{z}_{i \rightarrow j}^{k+1} - \mathbf{b}_{i}^{k+1} \, , 
\end{align*}
\normalsize
where $k$ denotes the current iteration; $\mathbf{u}_{i \rightarrow j}$ denotes the scaled dual variables for $\mathbf{z}_{i \rightarrow j}$; 
$Z^{k+1}_{\cdot \rightarrow \cdot}$ denotes the set of variables $\{ \mathbf{z}_{i \rightarrow j} \, | \, i \in \{1,\ldots d\}, j \in \{1, \ldots, d_i\}\}$.\footnote{Note that our formulation is a scaled ADMM, see e.g. \citep{boyd2011distributed} in Section 3.1.1. Therefore we have $\mathbf{u}_{i \rightarrow j}^{k+1} := \mathbf{u}_{i \rightarrow j}^{k} + \mathbf{z}_{i \rightarrow j}^{k+1} - \mathbf{b}_{i}^{k+1}$ and not $\mathbf{u}_{i \rightarrow j}^{k+1} := \mathbf{u}_{i \rightarrow j}^{k} + \rho (\mathbf{z}_{i \rightarrow j}^{k+1} - \mathbf{b}_{i}^{k+1})$.} Furthermore, we defined 
\begin{align*} 
g(B, \boldsymbol{\beta}_0) := 
&- \sum_{s = 1}^n \log \tilde{f}(y_s | \mathbf{x}_s, B, \boldsymbol{\beta}_0) \\
&+ \frac{\rho}{2} \sum_{i = 1}^{d} \sum_{j = 1}^{d_i} ||\mathbf{z}_{i \rightarrow e_i(j)}^k - \mathbf{b}_{i} +  \mathbf{u}_{i \rightarrow e_i(j)}^k ||_2^2  \, .
\end{align*}

\paragraph{Update of primal variables}
The update of $B$ can be solved with an approximate gradient Newton method \citep{byrd1995limited}, where fast calculation of the gradient and function evaluation is key to an efficient implementation.

\normalsize
Due to the second term, the calculation of $g(B, \boldsymbol{\beta}_0)$ is in $O(d \cdot \max_i d_i)$, and therefore, $O(d^2)$ for dense graphs. However, the double sum in the second term can be expressed as follows:\footnote{Details in Appendix \ref{sec:Appendix}.}
\begin{align*} 
&  \sum_{i = 1}^{d} \sum_{j = 1}^{d_i} ||\mathbf{z}_{i \rightarrow e_i(j)}^k - \mathbf{b}_{i} +  \mathbf{u}_{i \rightarrow e_i(j)}^k ||_2^2  \\
& = q_{all} 
- 2 \sum_{i = 1}^{d} \mathbf{b}_{i}^T \mathbf{q}\uparrow_i 
+ \sum_{i = 1}^{d} d_i \Big( \mathbf{b}_{i}^T \mathbf{b}_{i} \Big)  \, ,
\end{align*}
\normalsize
where we defined $\mathbf{q}_{ij} := \mathbf{z}_{i \rightarrow e_i(j)}^k +  \mathbf{u}_{i \rightarrow e_i(j)}^k$,
and $q_{all} := \sum_{i = 1}^{d} \sum_{j = 1}^{d_i} \mathbf{q}_{ij}^T \mathbf{q}_{ij}$, and 
$\mathbf{q}\uparrow_i :=  \sum_{j = 1}^{d_i} \mathbf{q}_{ij}$.
Since $q_{all}$ and $\mathbf{q}\uparrow_i$ can be precalculated, each repeated calculation of $g(B, \boldsymbol{\beta}_0)$ is in $O(d)$.

Furthermore, we note that 
\begin{align*}
\nabla_{\mathbf{b}_i} g(B, \boldsymbol{\beta}_0) 
&=  - \sum_{s = 1}^n \nabla_{\mathbf{b}_i} \log \tilde{f}(y_s | \mathbf{x}_s, B, \boldsymbol{\beta}_0)
- \rho \big( \mathbf{q}\uparrow_i - d_i \mathbf{b}_{i} \big) \, .
\end{align*}

\paragraph{Update of auxiliary variables}
The update of $\mathbf{z}_{i \rightarrow j}$ and $\mathbf{z}_{j \rightarrow i}$, for each unordered pair $\{i, j\}$ can be performed independently, i.e.:
\begin{align*}
& \forall \{i, j\} : i \in \{1,\ldots d\}, j \in \{1, \ldots, d_i\}: \\
& \mathbf{z}_{i \rightarrow j}^{k+1}, \mathbf{z}_{j \rightarrow i}^{k+1} 
%
:= \argmin_{\mathbf{z}_{i \rightarrow j}, \mathbf{z}_{j \rightarrow i}} \; \; 
\nu S_{i, j} ||\mathbf{z}_{i \rightarrow j} - \mathbf{z}_{j \rightarrow i}||_2 + \frac{\rho}{2} ||\mathbf{z}_{i \rightarrow j} - \mathbf{b}_{i}^{k+1} +  \mathbf{u}_{i \rightarrow j}^k ||_2^2 \\
&\qquad \qquad \qquad \qquad \qquad + \frac{\rho}{2} ||\mathbf{z}_{j \rightarrow i} - \mathbf{b}_{j}^{k+1} +  \mathbf{u}_{j \rightarrow i}^k ||_2^2 \, .
\end{align*} 
\normalsize
This optimization problem has a closed form solution, which was proven in a different context in \citep{hallac2015network}, with 
\begin{align*}
& \mathbf{z}_{i \rightarrow j}^{k+1} := \theta (\mathbf{b}_{i}^{k+1} - \mathbf{u}_{i \rightarrow j}^{k}) + (1 - \theta) (\mathbf{b}_{j}^{k+1} - \mathbf{u}_{j \rightarrow i}^k) \, , \\
& \mathbf{z}_{j \rightarrow i}^{k+1} :=  (1 - \theta) (\mathbf{b}_{i}^{k+1} - \mathbf{u}_{i \rightarrow j}^{k}) + \theta (\mathbf{b}_{j}^{k+1} - \mathbf{u}_{j \rightarrow i}^k) \, , 
\end{align*}
\normalsize
where 
\begin{align} \label{eq:edgeThresholdingEquation}
\theta := 
\begin{cases}
0.5 & \text{if } m = 0,\\
\max \Big(1 - \frac{\nu S_{i, j}}{\rho h } \, , \, 0.5 \Big)   
& \text{otherwise. }
\end{cases}
\end{align}
\normalsize
with $h :=  || \mathbf{b}_{i}^{k+1} - \mathbf{u}_{i \rightarrow j}^{k} - \mathbf{b}_{j}^{k+1} + \mathbf{u}_{j \rightarrow i}^k ||_2$.

\subsection{Identifying the covariate clusters}

Although in theory the optimization problem ensures that certain columns of $B$ are exactly $\mathbf{0}$, due to the use of ADMM this is not true in practice.\footnote{Details of the stopping criteria for ADMM are provided in Appendix \ref{sec:Appendix}.}
We therefore follow the strategy proposed in \citep{chi2015splitting}:
after convergence of the ADMM, we investigate $Z$, and place an edge between node $i$ and $j$, iff $\mathbf{z}_{i \rightarrow j}$ equals $\mathbf{z}_{j \rightarrow i}$.
Note that 
we can test this equality exactly (without numerical difficulties) due to the thresholding of $\theta$ to 0.5 in Equation \eqref{eq:edgeThresholdingEquation}.

In the resulting graph, there are two possible ways to identify clusters:
\begin{itemize}
	\item identify the connected components as clusters.\footnote{A component is in general not fully connected.} 
	\item consider only fully connected components as clusters.
\end{itemize}

Of course, in theory, since the optimization problem is convex, after complete convergence, we must have that the two ways result into the same clustering.
However, in practice, we find that the latter leads to too many covariates not being clustered (i.e. each covariate is in a single cluster). The latter is also computationally difficult, since identifying the largest fully connected component is NP-hard. Therefore, we proceed here with the former strategy.

We denote the identified clusters as $\mathcal{C}_1, \ldots, \mathcal{C}_m$, where $m$ is the number of clusters and 
$\mathscr{C} := \{ \mathcal{C}_1, \ldots, \mathcal{C}_m \}$ is a partition of the set of covariates.

\section{Approximate Bayesian Model Selection} \label{sec:approxBMS}

Note that different hyper-parameter settings for $\nu$ will result in different clusterings. 
For our experiments, we consider $\nu \in \{ n \cdot 2^{-0.1a} \, | \, a \in 0,1,2,\ldots, 299\}$.
This way, we get a range of clusterings. 
Like convex clustering, this allows us to plot a clustering hierarchy. However, in most situations, we are interested in finding the most plausible clustering, or ranking the clusterings according to some criterion.

One obvious way is to use cross-validation on the training data to estimate held-out classification accuracy for each setting of $\nu$.
However, the computational costs are enormous. 
Another, more subtle issue is that the group lasso terms in Equation \eqref{eq:convexSimpleFormulation} jointly perform clustering and shrinkage, but controlled by only one parameter $\nu$. 
However, the optimal clustering and the optimal shrinkage might not be achieved by the same value of $\nu$, an issue well known for the lasso penalty \citep{meinshausen2007relaxed}.
Therefore, we consider here a different model selection strategy: an approximate Bayesian model selection with low computational costs.

After we have found a clustering, we train a new logistic regression classifier with the parameter space for $B$ limited to the clustering.
Assuming some prior over $B$, the marginal likelihood $p(\mathbf{y})$ provides a trade-off between model fit and number of parameters (= number of clusters). 
A popular choice is the Bayesian information criterion (BIC)  \citep{schwarz1978estimating}, which assumes that the prior can be asymptotically ignored\footnote{That is, the prior $\log p(B)$ does not increase with $n$.}. However, BIC requires that the unpenalized maximum-likelihood estimate is defined. Unfortunately, this is not the case for logistic regression where linearly separable data will lead to infinite weights. 
For this reason, we suggest here to use a Gaussian prior on $B$ and then estimate the marginal likelihood with a Laplace approximation. 

In detail, in order to evaluate the quality of a clustering, we suggest to use the marginal likelihood $p(\mathbf{y} | X, \mathscr{C}, \boldsymbol{\beta}_0)$, where we treat the intercept vector $\boldsymbol{\beta}_0$ as hyper-parameter.
$X \in \mathbb{R}^{n \times d}$ denotes the design matrix, and $\mathbf{y} \in \{1,\ldots,c\}^n$ the responses. 

We define the following Bayesian model
\begin{align} \label{eq:simplifiedBayesModel}
p(\mathbf{y}, B | X, \mathscr{C}, \boldsymbol{\beta}_0) = \pi(B) \prod_s f(y_s | \mathbf{x}_{s}^{\mathscr{C}}, B, \boldsymbol{\beta}_0)
\end{align}
where $\pi$ denotes our prior on $B$, and $\mathbf{x}^{\mathscr{C}}$ denotes the projection of covariates $\mathbf{x}$ on the clustering defined by $\mathscr{C}$. This means
\begin{align*} 
\mathbf{x}^{\mathscr{C}} = T^{\mathscr{C}} \mathbf{x} \, ,
\end{align*}
where we define the matrix $T^{\mathscr{C}} \in \mathbb{R}^{m \times d}$ as follows
\begin{align*} 
T^{\mathscr{C}}_{ij} =
\left\{
\begin{array}{rl}
1 & \text{if } j \in \mathcal{C}_i \, , \\
0 & \text{else }  \, .
\end{array} \right.
\end{align*}

Assuming a Gaussian prior $N(0, \sigma^2)$ on each entry of $B \in \mathbb{R}^{m \times m}$, the log joint distribution for one data point $(y, \mathbf{x})$ is given by 
\begin{align*} 
&\log f(y | \mathbf{x} , B, \boldsymbol{\beta}_0) + \frac{1}{n} \log \pi(B) = \\
&\quad B_{y, \cdot} \mathbf{x} + \boldsymbol{\beta}_0(y) - \log \Big( \sum_{y'} \text{exp}(B_{y', \cdot} \mathbf{x}  + \boldsymbol{\beta}_0(y')) \Big)  \\
&\quad - \frac{1}{n} (\frac{1}{2 \sigma^2} ||B||_F^2 + \frac{cm}{2} \log (2 \pi \sigma^2) ) \, .
\end{align*}
\normalsize
For simplicity, let us denote $\mathbf{r}_i := (B_{i,\cdot})^T$. Note that $\mathbf{r}_i \in \mathbb{R}^m$.

Let us denote by $S(B)$ the Hessian of the log joint probability of the observed data $X = \{\mathbf{x}_1, \ldots, \mathbf{x}_n\}$ and prior $B$, i.e.
\begin{align*}
S(B)_{ij} := \sum_{s =1}^{n} \nabla_{\mathbf{r}_i, \mathbf{r}_j}^2 \{  \log f(y_s | \mathbf{x}_s, B, \boldsymbol{\beta}_0) + \frac{1}{n} \log \pi(B) \} \, ,
\end{align*}
where $S(B)_{ij} \in \mathbb{R}^{m \times m}$ denotes the $i,j$-th block of $S(B) \in \mathbb{R}^{cm \times cm}$.
By the Bayesian central limit theorem, we then have that the posterior is approximately normal with covariance matrix $- S(\hat{B})^{-1}$.
Then applying the Laplace approximation (see e.g. \citet{ando2010bayesian}), we get the following approximation for the marginal likelihood $p(X | \boldsymbol{\beta}_0)$:
\begin{align*}
p(X | \boldsymbol{\beta}_0) 
&= (2\pi)^{\frac{cm}{2}} |-S(\hat{B})|^{-\frac{1}{2}} \pi(\hat{B}) \prod_{s =1}^{n} f(y_s | \mathbf{x}_s , \hat{B}, \hat{\boldsymbol{\beta}}_0) \, ,
\end{align*}
where $\hat{B}$ and $\hat{\boldsymbol{\beta}}_0$ are the maximum a-posteriori (MAP) estimates from model \eqref{eq:simplifiedBayesModel}.
For large number of clusters $m$ the calculation of $|-S(\hat{B})^{-1}|$ is computationally expensive. Instead, we suggest to use only the diagonal of the Hessian.\footnote{The diagonal of the Hessian can be calculated for one sample 
	$(y, \mathbf{x})$ as follows
	$\text{Diag}(\nabla_{\mathbf{r}_i, \mathbf{r}_j}^2  \{ \log f(y | \mathbf{x}, B, \boldsymbol{\beta}_0) + \frac{1}{n} \log \pi(B) \})_u = 
	(p_i - 1) p_i \cdot x_z^2 - \frac{1}{\sigma^2 n} \, ,$
	with $i := \lceil \frac{u}{m} \rceil$, and $z := u - (i - 1)m$.}

We consider the intercept terms $\boldsymbol{\beta}_0$ as hyper-parameters and use empirical Bayes to estimate them, i.e. we set them to $\hat{\boldsymbol{\beta}}_0$.
The hyper-parameter $\sigma$ cannot be estimated with empirical Bayes, since the unpenalized maximum-likelihood might not be defined, and this would lead to $\sigma \rightarrow \infty$. Therefore, we estimate $\sigma$ using cross-validation on the full model, i.e. no clustering. 
This is computationally feasible since the MAP is an ordinary logistic regression with $\ell_2$ penalty and cross-validation needs to be done only \emph{once} and not for every clustering.

\section{Synthetic Data Experiments} \label{sec:experimentsSyntheticData}

Here, in this section, we investigate the performance of the proposed covariate clustering method, as well as the proposed model selection criterion on synthetic data. 

For all synthetic datasets, we set the number of classes $c$ to 4. The intercept vector $\boldsymbol{\beta}_0$ is set to the zero vector.
We group the covariates evenly into 10 clusters.
The weight vector $\mathbf{b}_i$ for covariate $i$ is set to the all zero vector except one position which is set to $0.5 \cdot \text{ground truth cluster id}$.
That means each covariate is associated with exactly one class. 
If two covariates $i$ and $j$ belong to the same cluster (ground truth), then $\mathbf{b}_i$ and $\mathbf{b}_j$ are equal. 
Finally, we generate samples from a multivariate normal distribution as follows:
given a positive definite covariate similarity matrix $S^*$, we generate a sample $\mathbf{x}$ from class $y$ by $\mathbf{x} \sim N(B_{y,\cdot}, S^*)$.
For each class we generate the same number of samples.
We consider $d \in \{40, 200, 1000\}$ and $n \in \{40, 400, 4000\}$.

\begin{figure}[h]
	\centering
	\includegraphics[page=2,scale=0.45,trim=15cm 12cm 20cm 0cm]{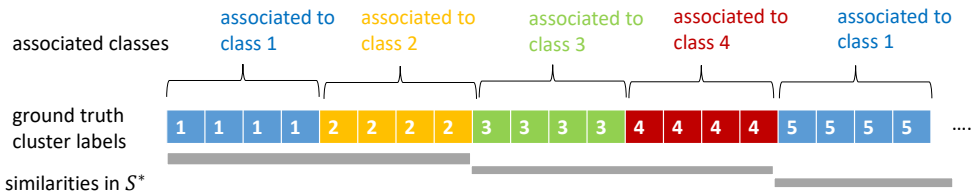} 
	\caption{Shows the first 20 covariates with their associated cluster labels (ground truth), associated classes, and similarities between the covariates. Grey bar at the same height means that the covariates are similar to each other, i.e. $S^*_{ij} = 0.9$, otherwise, for non-similar covariate $i$ and $j$, we set $S^*_{ij} = 0.0$.
		\label{fig:syntheticDataContradicts}}
\end{figure}

In practice, the a-priori similarity information between covariates will not fully agree with the associated classes.
For that reason, we define $S^*$ such that the cluster structure implied by $S^*$ 
covers covariates that are associated to two different class labels.
An example is shown in Figure \ref{fig:syntheticDataContradicts}.


\paragraph{Estimation of $S$}
Since in practice, we only have access to a noisy estimate of $S^*$, 
rather than setting $S:= S^*$, we estimate $S$ from the data, by first normalizing the data to have zero mean for each class, and then pooling all samples from all classes for acquiring an estimate of the covariance matrix $\hat{\Sigma}$ using the Ledoit-Wolf shrinkage estimator \citep{ledoit2004well}.
Finally, we hard-threshold all negative entries in $\hat{\Sigma}$ to acquire $S$.

\paragraph{Baseline Methods}

\begin{table}
	\caption{Shows the ANMI score of the proposed method, convex clustering, and $k$-means clustering on synthetic data.
		Standard deviation in brackets. 
		For hyper-parameter selection the proposed marginal likelihood criteria is used. Small number shows the ANMI score when cross-validation is used instead of the marginal likelihood.}
	\label{tab:resultsSynData_ANMI}
	\centering
	\footnotesize
	\begin{tabular}{ccccc}
		\\
		\multicolumn{5}{c}{Proposed Model} \\ 
		\midrule 
		&  & \multicolumn{3}{c}{n} \\ 
		&  &  \bf 40 & \bf 400 & \bf 4000 \\ 
		\multirow{3}{*}{d} & \bf 40 & 0.84 (0.11)  {\tiny 0.89 (0.03) } & 0.71 (0.2)  {\tiny 0.9 (0.0) } & 1.0 (0.0)  \\ 
		& \bf 200 & 0.87 (0.16)  {\tiny 0.87 (0.13) } & 0.95 (0.03)  {\tiny 0.93 (0.04) } & 1.0 (0.0)  \\ 
		& \bf 1000 & 0.94 (0.0)  & 0.94 (0.0)  & 1.0 (0.0)  \\ 
		\midrule
		\multicolumn{5}{c}{Convex Clustering} \\
		\midrule
		&  & \multicolumn{3}{c}{n} \\ 
		&  &  \bf 40 & \bf 400 & \bf 4000 \\ 
		\multirow{3}{*}{d} & \bf 40 & 0.59 (0.04)  {\tiny 0.58 (0.05) } & 0.54 (0.18)  {\tiny 0.48 (0.23) } & 0.38 (0.27)  \\ 
		& \bf 200 & 0.69 (0.0)  {\tiny 0.69 (0.0) } & 0.69 (0.0)  {\tiny 0.69 (0.0) } & 0.48 (0.31)  \\ 
		& \bf 1000 & 0.01 (0.01)  & 0.01 (0.0)  & 0.0 (0.0)  \\ 
		\midrule
		\multicolumn{5}{c}{$k$-means Clustering} \\ 
		\midrule 
		&  & \multicolumn{3}{c}{n} \\ 
		&  &  \bf 40 & \bf 400 & \bf 4000 \\ 
		\multirow{3}{*}{d} & \bf 40 & 0.55 (0.09)  {\tiny 0.52 (0.12) } & 0.43 (0.14)  {\tiny 0.43 (0.14) } & 0.35 (0.21)  \\ 
		& \bf 200 & 0.56 (0.2)  {\tiny 0.58 (0.09) } & 0.57 (0.15)  {\tiny 0.53 (0.14) } & 0.54 (0.13)  \\ 
		& \bf 1000 & 0.65 (0.06)  & 0.7 (0.0)  & 0.41 (0.08)  \\
		\bottomrule
	\end{tabular}
\end{table}

We compare the proposed method to $k$-means with the initialization proposed in \citep{arthur2007k} and convex clustering \citep{chi2015splitting}, which both use the same similarity matrix $S$ for clustering.
For $k$-means, we consider $k \in \{1,2, \ldots, d\}$. For convex clustering the hyper-parameter $\gamma$ is tested in the range 0 to 5 evenly spaced with step size $1/300$, and weights $w_{ij}$ are set to the exponential kernel with $\phi = 0.5$.\footnote{For $d > 1000$ the convex clustering method led to memory overflow. Therefore, we needed to limit the weights to 5-nearest neighbors.}
After clustering with then train an ordinary multinomial logistic regression model.
\footnote{On the other hand, first classification and then covariate clustering makes it difficult to incorporate the similarity information in $S$.}

\paragraph{Clustering Evaluation}
We repeat each experiment 10 times and report average and standard deviation of the adjusted normalized mutual information (ANMI)  \citep{vinh2010information} when compared to the true clustering. 
The ANMI score ranges from 0.0 (agreement with true clustering at pure chance level) to 1.0 (complete agreement with true clustering).
All results are summarized in Table \ref{tab:resultsSynData_ANMI}.
For selecting a clustering we use the approximate marginal likelihood selection criterion described in Section \ref{sec:approxBMS}, we also show the results when using cross-validation\footnote{We use 5-fold cross-validation selecting the smallest model that is within one standard deviation from the best model, as suggested in \citep{hastie2009elements}.} (small font size), where it is computationally feasible.

\paragraph{Quality of marginal likelihood model selection criterion}
Comparing the results when using cross-validation with the proposed approximate marginal likelihood criteria, we see that in most cases, there is no significant difference. However, compared to $k$-fold cross-validation, the approximate marginal likelihood criteria is $k$ times more computationally efficient. 

\paragraph{Quality of clustering result}
Even for small number of samples, $n \leq 400$, the proposed method's clustering result is close to the ground truth, and fully recovers the correct clustering given sufficient number of samples, $n = 4000$. This is neither the case for convex clustering, nor $k$-means clustering. Moreover, we find that for large number of variables, $d = 1000$, convex clustering performed poorly.
Finally, we note that in the more unrealistic scenario, where the a-priori similarity information fully agrees with the class association, convex clustering and $k$-means perform similar to the proposed method (see results in Appendix \ref{sec:Appendix}).

\subsection{Runtime Experiments} \label{subSec:runtimeExperiments}

In order to check the efficiency of our proposed ADMM solution, we compare it to two standard solvers for convex optimization, namely ECOS \citep{bib:Domahidi2013ecos} and SCS \citep{o2016conic} using the CVXPY interface \citep{cvxpy,cvxpy_rewriting}. Note that SCS uses a generic ADMM method for which we use the same stopping criteria as our method. 
We run all experiments on a cluster with 88 Intel(R) Xeon(R) CPUs with 2.20GHz.
%
We repeated each experiment 10 times and report the average runtime (wall-clock time) in
in Table \ref{tab:resultsSynDataRuntime}. 

Our evaluation shows that the proposed ADMM scales well, both in terms of number of samples $n$, and in terms of number of variables $d$. Moreover, the proposed ADMM is considerably faster than the standard solvers when $n$ or $d$ are large, enabling us to  apply the proposed method to real data.


\begin{table}
	\caption{Average runtime (in minutes with standard deviation in brackets) of the proposed ADMM solution, and two standard solvers ECOS \citep{bib:Domahidi2013ecos} and SCS \citep{o2016conic}. When a method was too slow to perform 10 repeated experiments, we either ran only 3 experiments, or estimated the expected runtime based on a few iterations, marked as $(*)$.} 
	\label{tab:resultsSynDataRuntime}
	\centering
	\footnotesize
	\begin{tabular}{ccccc}
		\\
		\multicolumn{5}{c}{Solving with proposed ADMM} \\ 
		\midrule 
		&  & \multicolumn{3}{c}{n} \\ 
		&  &  \bf 40 & \bf 400 & \bf 4000 \\ 
		\multirow{3}{*}{d} & \bf 40 & 5.314 (0.503)  &  12.271 (0.488)  & 63.318 (1.339)  \\ 
		& \bf 200 & 19.331 (4.031)  & 86.815 (4.246)  &   187.734 (4.101)  \\ 
		& \bf 1000 & 102.721 (7.896) & 552.763 (51.104)  &  982.877 (12.316)  \\ 
		\midrule
		\multicolumn{5}{c}{Solving with ECOS} \\
		\midrule
		&  & \multicolumn{3}{c}{n} \\ 
		&  &  \bf 40 & \bf 400 & \bf 4000 \\ 
		\multirow{3}{*}{d} & \bf 40 &  6.75 (0.841)  &  18.936 (0.186)   &   360.677 (11.68)    \\ 
		& \bf 200 & 147.6 (35.325)    & 196.136 (8.52)  & 3800(*)   \\ 
		& \bf 1000 & 15400(*)\footnote{For this setting ECOS did not converge properly, which led to exceptional long runtime.}  & 6990(*)  & 65100(*)   \\ 
		\midrule
		\multicolumn{5}{c}{Solving with SCS} \\ 
		\midrule 
		&  & \multicolumn{3}{c}{n} \\ 
		&  &  \bf 40 & \bf 400 & \bf 4000 \\ 
		\multirow{3}{*}{d} & \bf 40 & 11.333 (1.159)   &  155.13 (5.767)   &  1590 (*)   \\ 
		& \bf 200 & 163.145(33.505)    & 360.96(20.762)   & 1680(*)   \\ 
		& \bf 1000 & 15300(*)  & 5730(*)  &  23100(*)   \\ 
		\bottomrule
	\end{tabular}
\end{table}

\section{Experiments on Real Data} \label{sec:experimentsRealData}

\begin{table}
	\caption{Results for Newsgroup20 and IMDB dataset. Shows the result of each clustering method, when hyper-parameters are selected with the approximate marginal likelihood.
		For reference, also the results when performing no clustering, i.e. the full model, are shown.}
	\label{tab:resultsNewsgroup20_and_IMDB}
	\centering
	\footnotesize
	\begin{tabular}{llll}
		\toprule 
		\multicolumn{4}{c}{Newsgroup20} \\
		\midrule 
		Method & Marginal Likelihood & Nr. Clusters & Accuracy \\
		\midrule
		Proposed   & \bf{-9487.1} & 1173 & 0.86 \\ 
		Convex  & -10695.8 & 1538 & \bf{0.88} \\ 
		$k$-means  & -10152.4 & \bf{1036} & 0.85 \\ 
		No Clustering  & -10695.8 & 1538 & \bf{0.88} \\ 
		\midrule 
		\multicolumn{4}{c}{IMDB} \\
		\midrule
		Proposed   & \bf{-578.4} & \bf{140} & \bf{0.85} \\ 
		Convex   & -3637.6 & 2604 & \bf{0.85} \\ 
		$k$-means  & -3589.7 & 2528 & \bf{0.85} \\ 
		No Clustering  & -3637.6 & 2604 & \bf{0.85} \\ 
		\bottomrule
	\end{tabular}
\end{table}

For our experiments on real data, we used the movie review corpus IMDB \citep{maas2011learning}, which consists of in total 100k (labeled and unlabeled) documents. IMDB is a balanced corpus with 50\% of the movies being reviewed as ``good movie", and the remaining as ``bad movie''.  
As the second dataset, we used the 20 Newsgroups corpus (Newsgroup20)\footnote{\url{http://people.csail.mit.edu/jrennie/20newsgroups/}} with around 19k documents categorized into 20 classes (topics like ``hockey'', ``computer'',...).
In order to check whether our model selection criterion correlates well with accuracy on held-out data, we use 10000 documents for training and clustering selection, and the remaining documents as held-out data.\footnote{39k and 9k held-out documents for IMDB and Newsgroup20, respectively.}
We removed all duplicate documents, and performed tokenization and stemming with Senna \citep{collobert2011natural}. Furthermore, we removed irrelevant covariates (details in Appendix \ref{sec:Appendix}), leading to 2604 and 1538 covariates for IMDB and Newsgroup20, respectively.

\subsection{Covariate Similarity Measure}
From additional unlabeled documents, we determined the similarity between two covariates $i$ and $j$ as follows.

First, using the unlabeled datasets, we created for each covariate $i$ a word embedding $\mathbf{w}_i$.
For IMDB, we created 50-dimensional word embeddings with word2vec \citep{mikolov2013distributed} using 75k documents from the IMDB corpus.\footnote{\url{https://code.google.com/p/word2vec/} word2vec was used with the default settings.}
For Newsgroup20, since, the number of samples is rather small, we used the 300 dimensional word embeddings from GloVe \citep{pennington2014glove} that were trained on Wikipedia + Gigaword 5.
Finally, the similarity between covariate $i$ and $j$ was calculated using $S_{ij} = e^{- \frac{1}{2} ||\mathbf{w}_i - \mathbf{w}_j||_2^2 }$.
Using the similarity matrix $S$, the baseline methods are trained as in Section \ref{sec:experimentsSyntheticData}.

\subsection{Quantitative Results}

For real data, no ground truth for clustering is available. 
Therefore, we suggest to use the number of clusters vs the classification accuracy on held-out data. An ideal covariate clustering method should achieve high classification accuracy, while having only few clusters. Note that fewer number of clusters leads to more compact models, and potentially eases  interpretability. 
%

For selecting a clustering we use the proposed marginal likelihood criterion (Section \ref{sec:approxBMS}), and the held-out data is only used for final evaluation. 
The results for IMDB and Newsgroup20 are shown in Table \ref{tab:resultsNewsgroup20_and_IMDB}. 
We find that the marginal likelihood criterion tends to select models which accuracy on held-out data is similar to the full model, but with fewer number of parameters.
Furthermore, for IMDB, we find that the proposed method leads to considerably more compact models than convex clustering and $k$-means, while having similar held-out accuracy. On the other hand, for Newsgroup20, the accuracies of the proposed method and $k$-means clustering are similar, indicating that there is good agreement between the similarity measure $S$ and the classification task.

In order to confirm that these conclusions are true independently of the model selection criterion, we also show in Table \ref{tab:resultsNewsgroup20_and_IMDB_cmpToKMeansInDetail} the held-out accuracy of clusterings with number of clusters being around 100, 500 and 1000.\footnote{Note that, in contrast to $k$-means, the proposed method and convex clustering can only control the number of clusters indirectly through their regularization hyper-parameter.} The results confirm that the proposed method can lead to better covariate clusterings than $k$-means and convex clustering.

\begin{table}
	\caption{Results for Newsgroup20 and IMDB dataset. Comparison of held-out accuracy of the proposed method, convex clustering and $k$-means for around the same number of clusters: 100, 500, 1000. 
	}
	\label{tab:resultsNewsgroup20_and_IMDB_cmpToKMeansInDetail}
	\centering
	\footnotesize
	\begin{tabular}{llll}
		\toprule 
		\multicolumn{4}{c}{Newsgroup20} \\
		\midrule 
		Nr. Clusters & Proposed  & Convex & $k$-means \\
		\midrule
		100 & \bf{0.67} & 0.44 & \bf{0.67} \\ 
		500 & \bf{0.80} & 0.70 & 0.79 \\ 
		1000 & \bf{0.84} & 0.82 & 0.83 \\ 
		\midrule 
		\multicolumn{4}{c}{IMDB} \\
		\midrule 
		Nr. Clusters & Proposed  & Convex & $k$-means \\
		\midrule
		100 & \bf{0.85} & 0.59 & 0.8 \\ 
		500 & \bf{0.84} & 0.76 & 0.82 \\ 
		1000 & \bf{0.84} & 0.8 & 0.82 \\ 
		\bottomrule
	\end{tabular}
\end{table}

\subsection{Qualitative Results}

Like convex clustering, our proposed method can be used to create a hierarchical clustering structure by varying the hyper-parameter $\nu$. In particular, for large enough $\nu$, all covariates collapse to one cluster. On the other side, for $\nu = 0$, each covariate is assigned to a separate cluster. Visualizations of the results from the proposed method, convex clustering and $k$-means are given in Appendix \ref{sec:Appendix}.
As expected, we observe that $k$-means and convex clustering tend to produce clusters that include covariates that are associated to different classes.
For example, in IMDB, with $k$-means and convex clustering, all number ratings are clustered together. However, these have very different meaning, since for example 7/10, 8/10, 9/10, 10/10 rate good movies and 0, 1/10, *1/, 2/10,... rate bad movies. 
In contrast, our proposed method distinguishes them correctly by assigning them to different clusters associated with different classes.

\section{Conclusions} \label{sec:conclusions}

We presented a new method for covariate clustering that uses an \emph{a-priori} similarity measure between two covariates, and additionally, class label information. 
In contrast to $k$-means and convex clustering, the proposed method creates clusters that are \emph{a-posteriori} plausible in the sense that they help to explain the observed data (samples with class label information).
Like ordinary convex clustering \citep{chi2015splitting}, the proposed objective function is convex, and therefore insensitive to heuristics for initializations (as needed by $k$-means clustering).

Solving the convex objective function is computationally challenging. Therefore, we proposed an efficient ADMM algorithm which allows us to scale to several 1000 variables. 
Furthermore, in order to prevent computationally expensive cross-validation, we proposed a marginal likelihood criterion similar to BIC.
For synthetic and real data, we confirmed the usefulness of the proposed clustering method and the marginal likelihood criterion.

\appendix
\section{Appendices} \label{sec:Appendix}
\subsection{Detailed Derivations of ADMM Update Equations}

\paragraph{Update of primal variables}
The double sum in the second term of $g(B, \boldsymbol{\beta}_0)$ can be expressed as follows:
\begin{align*} 
&  \sum_{i = 1}^{d} \sum_{j = 1}^{d_i} ||\mathbf{z}_{i \rightarrow e_i(j)}^k - \mathbf{b}_{i} +  \mathbf{u}_{i \rightarrow e_i(j)}^k ||_2^2   \\
& = \sum_{i = 1}^{d} \sum_{j = 1}^{d_i} ||\mathbf{q}_{ij} - \mathbf{b}_{i} ||_2^2 \\
& = \sum_{i = 1}^{d} \sum_{j = 1}^{d_i} \Big( \mathbf{q}_{ij}^T \mathbf{q}_{ij} - 2 \cdot \mathbf{b}_{i}^T\mathbf{q}_{ij} + \mathbf{b}_{i}^T \mathbf{b}_{i} \Big) \\ 
& = \sum_{i = 1}^{d} \sum_{j = 1}^{d_i} \Big( \mathbf{q}_{ij}^T \mathbf{q}_{ij} \Big) - 2 \sum_{i = 1}^{d} \mathbf{b}_{i}^T \Big( \sum_{j = 1}^{d_i} \mathbf{q}_{ij} \Big) + \sum_{i = 1}^{d} \sum_{j = 1}^{d_i} \Big( \mathbf{b}_{i}^T \mathbf{b}_{i} \Big) \\ 
& = q_{all} 
- 2 \sum_{i = 1}^{d} \mathbf{b}_{i}^T \mathbf{q}\uparrow_i 
+ \sum_{i = 1}^{d} d_i \Big( \mathbf{b}_{i}^T \mathbf{b}_{i} \Big)  \, ,
\end{align*}
\normalsize
where we defined $\mathbf{q}_{ij} := \mathbf{z}_{i \rightarrow e_i(j)}^k +  \mathbf{u}_{i \rightarrow e_i(j)}^k$,
and $q_{all} := \sum_{i = 1}^{d} \sum_{j = 1}^{d_i} \mathbf{q}_{ij}^T \mathbf{q}_{ij}$, and 
$\mathbf{q}\uparrow_i :=  \sum_{j = 1}^{d_i} \mathbf{q}_{ij}$.
Since $q_{all}$ and $\mathbf{q}\uparrow_i$ can be precalculated, each repeated calculation of $g(B, \boldsymbol{\beta}_0)$ is in $O(d)$.

Furthermore, we note that 
\begin{align*}
\nabla_{\mathbf{b}_i} g(B, \boldsymbol{\beta}_0) &=  - \sum_{s = 1}^n \nabla_{\mathbf{b}_i} \log \tilde{f}(y_s | \mathbf{x}_s, B, \boldsymbol{\beta}_0) \\
&\quad + \rho \sum_{j = 1}^{d_i} \Big( - \mathbf{z}_{i \rightarrow e_i(j)}^k + \mathbf{b}_{i} -  \mathbf{u}_{i \rightarrow e_i(j)}^k \Big) \\
&=  - \sum_{s = 1}^n \nabla_{\mathbf{b}_i} \log \tilde{f}(y_s | \mathbf{x}_s, B, \boldsymbol{\beta}_0)
- \rho \sum_{j = 1}^{d_i}\mathbf{q}_{ij} + \rho \sum_{j = 1}^{d_i} \mathbf{b}_{i} \\
&=  - \sum_{s = 1}^n \nabla_{\mathbf{b}_i} \log \tilde{f}(y_s | \mathbf{x}_s, B, \boldsymbol{\beta}_0)
- \rho \big( \mathbf{q}\uparrow_i - d_i \mathbf{b}_{i} \big) \, .
\end{align*}
\normalsize

\paragraph{Update of auxiliary variables}
The update of $\mathbf{z}_{i \rightarrow j}$ and $\mathbf{z}_{j \rightarrow i}$, for each unordered pair $\{i, j\}$ can be performed independently, i.e.:
\begin{align*}
& \forall \{i, j\} : i \in \{1,\ldots d\}, j \in \{1, \ldots, d_i\}: \\
& \mathbf{z}_{i \rightarrow j}^{k+1}, \mathbf{z}_{j \rightarrow i}^{k+1} 
:= \argmin_{\mathbf{z}_{i \rightarrow j}, \mathbf{z}_{j \rightarrow i}} \; \; 
\nu \frac{1}{2} (S_{i, j} ||\mathbf{z}_{i \rightarrow j} - \mathbf{z}_{j \rightarrow i}||_2 + S_{j, i} ||\mathbf{z}_{j \rightarrow i} - \mathbf{z}_{i \rightarrow j}||_2) \\
& \qquad \qquad \qquad \qquad \; \; + \frac{\rho}{2} ||\mathbf{z}_{i \rightarrow j} - \mathbf{b}_{i}^{k+1} +  \mathbf{u}_{i \rightarrow j}^k ||_2^2 \\
& \qquad \qquad \qquad \qquad \; \; + \frac{\rho}{2} ||\mathbf{z}_{j \rightarrow i} - \mathbf{b}_{j}^{k+1} +  \mathbf{u}_{j \rightarrow i}^k ||_2^2\\
& \qquad \qquad \; 
= \argmin_{\mathbf{z}_{i \rightarrow j}, \mathbf{z}_{j \rightarrow i}} \; \; 
\nu S_{i, j} ||\mathbf{z}_{i \rightarrow j} - \mathbf{z}_{j \rightarrow i}||_2 \\
&\qquad \qquad \qquad \qquad \; \; + \frac{\rho}{2} ||\mathbf{z}_{i \rightarrow j} - \mathbf{b}_{i}^{k+1} +  \mathbf{u}_{i \rightarrow j}^k ||_2^2 \\
&\qquad \qquad \qquad \qquad \; \; + \frac{\rho}{2} ||\mathbf{z}_{j \rightarrow i} - \mathbf{b}_{j}^{k+1} +  \mathbf{u}_{j \rightarrow i}^k ||_2^2 \, .
\end{align*}

\subsection{Stopping criteria for ADMM} \label{sec:stoppingCriteria}

As stopping criteria, we use, as suggested in \citep{boyd2011distributed}, the $\ell_2$ norm of the primal $\mathbf{r}^k$ and dual residual $\mathbf{s}^k$ at iteration $k$, defined here as \\
\begin{align*} 
& ||\mathbf{r}^k||_2^2 := \sum_{i = 1}^{d} \sum_{j = 1}^{d_i} || \mathbf{z}_{i \rightarrow e_i(j)}^k - \mathbf{b}_i^k ||_2^2 \, \\
& ||\mathbf{s}^k||_2^2 := \rho \sum_{i = 1}^{d} \sum_{j = 1}^{d_i} || \mathbf{z}_{i \rightarrow e_i(j)}^{k} -  \mathbf{z}_{i \rightarrow e_i(j)}^{k-1} ||_2^2 \, . 
\end{align*}
\normalsize
We stop after iteration $k$, if 
\begin{align*} 
& ||\mathbf{r}^k||_2 < \sqrt{c(d + 2 l)} \, \epsilon \, \text{, and } \, ||\mathbf{s}^k||_2 < \sqrt{c(d + 2 l)} \, \epsilon \, ,
\end{align*}
where $l$ is the total number of edges in our graph, and $\epsilon$ is set to 0.00001.
We also stop, in the cases where $k \geq 1000$.

\subsection{Details of Approximate Bayesian Model Selection}

The gradient with respect to $\mathbf{r}_i$ is
\begin{align*} 
& \nabla_{\mathbf{b}_i} \{ \log f(y | \mathbf{x}, B, \boldsymbol{\beta}_0) + \frac{1}{n} \log \pi(B) \} \\
& = 
\left\{
\begin{array}{rl}
- p_{i} \cdot \mathbf{x} - \frac{1}{\sigma^2 n} \mathbf{r}_i  & \text{if } i \neq y \, , \\
(1 - p_{i}) \cdot \mathbf{x} - \frac{1}{\sigma^2 n} \mathbf{r}_i  & \text{if } i = y \, ,
\end{array} \right.
\end{align*}
with $p_{i} := \frac{exp(\mathbf{r}_{i}^T \mathbf{x} + \beta_{i})}{\sum_{y'} exp(\mathbf{r}_{y'}^T \mathbf{x} + \beta_{y'})}$.

The Hessian with respect to $\mathbf{r}_i$,  $\mathbf{r}_j$ is
\begin{align*} 
& \nabla_{\mathbf{r}_i, \mathbf{r}_j}^2 \{  \log f(y | \mathbf{x}, B, \boldsymbol{\beta}_0) + \frac{1}{n} \log \pi(B) \} \\
& = 
\left\{
\begin{array}{rl}
p_i p_j \cdot \mathbf{x} \mathbf{x}^T & \text{if } i \neq j \, , \\
(p_i - 1) p_i \cdot \mathbf{x} \mathbf{x}^T - \frac{1}{\sigma^2 n} I_{d}  & \text{if } i = j \, .
\end{array} \right.
\end{align*}

\subsection{Additional Experiments on Synthetic Data}

\begin{figure}[h]
	\centering
	\includegraphics[page=1,scale=0.45,trim=15cm 12cm 20cm 0cm]{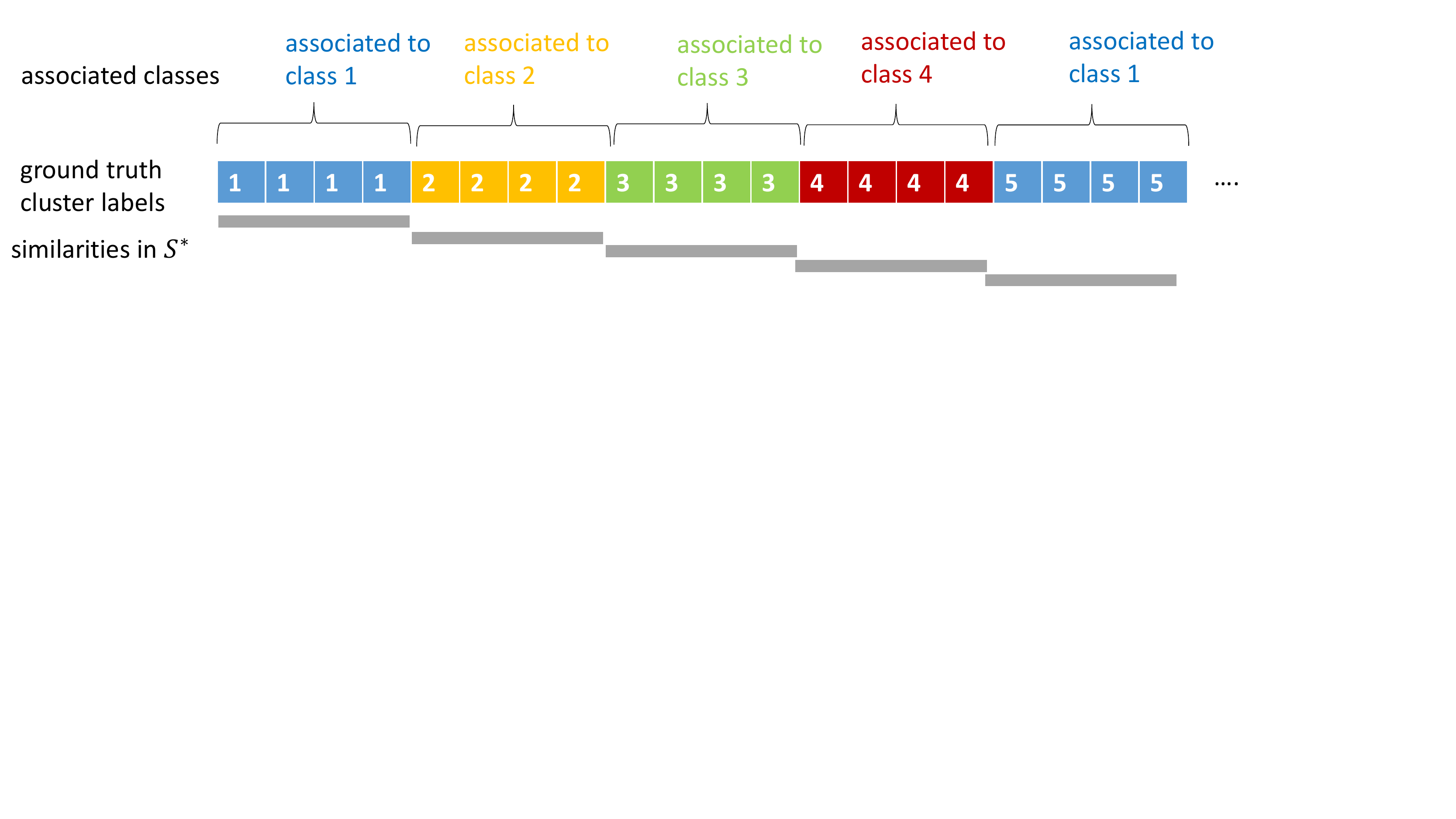} 
	\caption{Shows the first 20 covariates with their associated cluster labels (ground truth), associated classes, and similarities between the covariates. Grey bar at the same height means that the covariates are similar to each other, i.e. $S^*_{ij} = 0.9$. Here, similarity $S^*$ agrees with class information.
		\label{fig:syntheticDataAgrees}}
\end{figure}

\paragraph{$S^*$ agrees with class label information}
If covariates $i$ and $j$ are in different clusters, then $S^*_{ij} = 0$, otherwise $S^*_{ij} = 0.9$. $S^*_{ii}$ is set to $1.0$.
An example, where each cluster has four covariates is show in Figure \ref{fig:syntheticDataAgrees}.

\begin{table}
	\caption{Shows the ANMI score of the proposed method, convex clustering, and $k$-means clustering on synthetic data: prior covariate similarity is fully agreeing as shown in Figure \ref{fig:syntheticDataAgrees}. Standard deviation in brackets. 
		For hyper-parameter selection the proposed marginal likelihood criteria is used. Small number shows the ANMI score when cross-validation is used instead of the marginal likelihood.}
	\label{tab:resultsSynData_ANMI}
	\centering
	\footnotesize
	\begin{tabular}{ccccc}
		\\
		\multicolumn{5}{c}{Proposed Model} \\ 
		\midrule 
		&  & \multicolumn{3}{c}{n} \\ 
		&  &  \bf 40 & \bf 400 & \bf 4000 \\ 
		\multirow{3}{*}{d} & \bf 40 & 0.7 (0.12)  {\tiny 0.0 (0.0) } & 0.83 (0.16)  {\tiny 1.0 (0.0) } & 1.0 (0.0)  \\ 
		& \bf 200 & 0.88 (0.07)  {\tiny 1.0 (0.0) } & 0.92 (0.1)  {\tiny 1.0 (0.0) } & 1.0 (0.0)  \\ 
		& \bf 1000 & 0.95 (0.07)  & 0.96 (0.06)  & 1.0 (0.0)  \\ 
		\midrule
		\multicolumn{5}{c}{Convex Clustering} \\ 
		\midrule
		&  & \multicolumn{3}{c}{n} \\ 
		&  &  \bf 40 & \bf 400 & \bf 4000 \\ 
		\multirow{3}{*}{d} & \bf 40 & 0.93 (0.13)  {\tiny 0.93 (0.13) } & 1.0 (0.0)  {\tiny 1.0 (0.0) } & 1.0 (0.0)  \\ 
		& \bf 200 & 0.99 (0.04)  {\tiny 0.98 (0.04) } & 1.0 (0.0)  {\tiny 1.0 (0.0) } & 1.0 (0.0)  \\ 
		& \bf 1000 & 0.04 (0.02)  & 0.0 (0.0)  & 0.0 (0.0)  \\ 
		\midrule
		\multicolumn{5}{c}{$k$-means Clustering} \\ 
		\midrule 
		&  & \multicolumn{3}{c}{n} \\ 
		&  &  \bf 40 & \bf 400 & \bf 4000 \\  \multirow{3}{*}{d} & \bf 40 & 0.72 (0.21)  {\tiny 0.66 (0.22) } & 0.84 (0.2)  {\tiny 0.77 (0.25) } & 0.99 (0.03)  \\ 
		& \bf 200 & 0.92 (0.11)  {\tiny 0.88 (0.16) } & 0.93 (0.07)  {\tiny 0.88 (0.09) } & 0.99 (0.02)  \\ 
		& \bf 1000 & 0.98 (0.03)  & 1.0 (0.0)  & 0.95 (0.07)  \\ 
		\bottomrule
	\end{tabular}
\end{table}

\subsection{Additional Details and Visualization of Real Data Results} 

\subsubsection{Covariate Selection}
First, for each dataset we extract the 10000 most frequent words as covariates, and represent each document $s$ by a vector $\mathbf{x}_s$ where each dimension contains the tf-idf score of each word in the document.\footnote{We also perform $\ell_2$ scaling of each $\mathbf{x}_s$ which is known to improve classification performance. This is performed before the normalization of the covariates.} Finally, we normalize the covariates to have mean 0 and variance 1. 

For such high dimensional problems, many covariates are only negligibly relevant. Therefore, in order to further reduce the dimension, we apply a group lasso penalty on the columns of $B$ like in \citep{vincent2014sparse}\footnote{We use also an ADMM algorithm to perform this optimization. We omit the details since the optimization function is the same as in Definition 1 in \citep{vincent2014sparse} with $\alpha = 1$ and $\gamma \in \{ 2^{-a} \, | \, a \in 0,1,2,\ldots, 29\}$.} and select the model with the highest approximate marginal likelihood. 
This leaves us with 2604 and 1538 covariates for IMDB and Newsgroup20, respectively.

\subsubsection{Qualitative Results}

\begin{figure}[t]
	\centering
	\includegraphics[trim=0 440 0 7620,clip=true,scale=0.4]{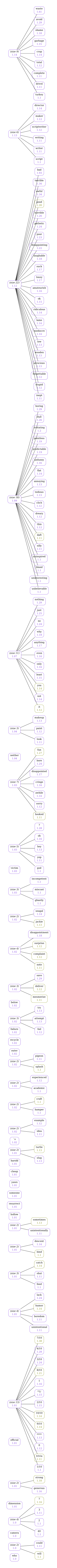}
	\caption{Shows part of the clustering result of $k$-means clustering on IMDB for the class bad movie. 
		\label{fig:IMDB_kMeansClustering_bad_movie}}
\end{figure}

\begin{figure}[t]
	\centering
	\includegraphics[trim=360 7140 4980 1820,clip=true,scale=0.4]{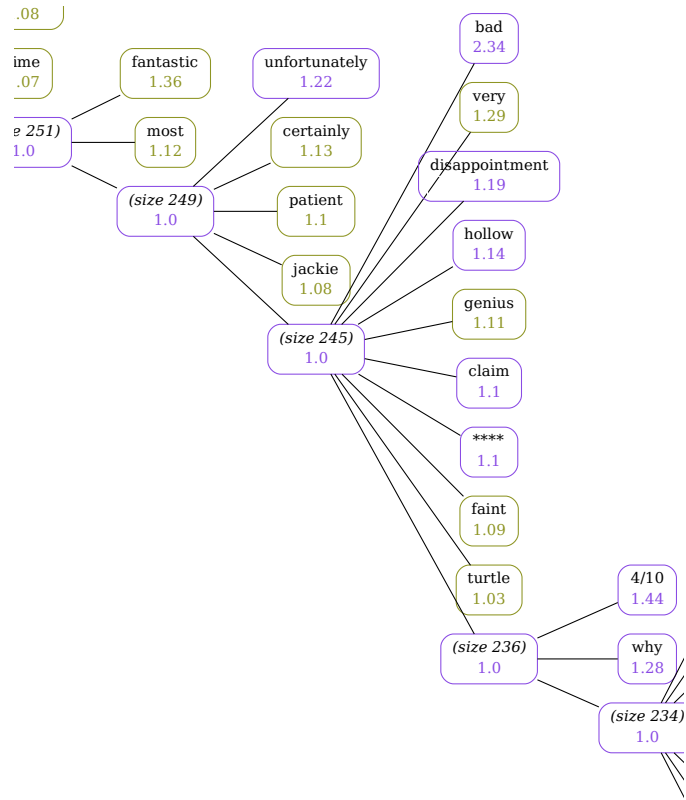}
	\caption{Shows part of the clustering result of convex clustering on IMDB for the class bad movie. 
		\label{fig:IMDB_convexClustering_bad_movie}}
\end{figure}

\begin{figure}[t]
	\centering
	\includegraphics[trim=0 3570 0 1030,clip=true,scale=0.4]{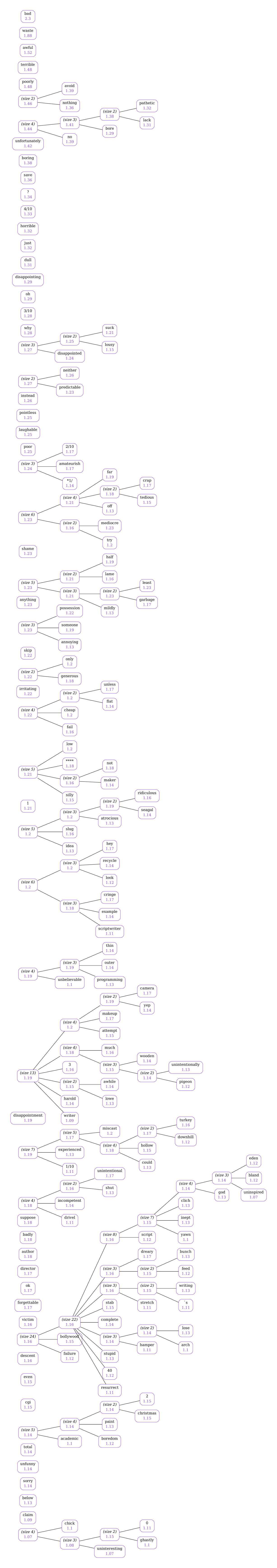}
	\caption{Shows part of the clustering result of the proposed method on IMDB for the class bad movie.
		\label{fig:IMDB_proposed_bad_movie}}
\end{figure}

\begin{figure}[t]
	\centering
	\includegraphics[trim=0 3820 0 30,clip=true,scale=0.35]{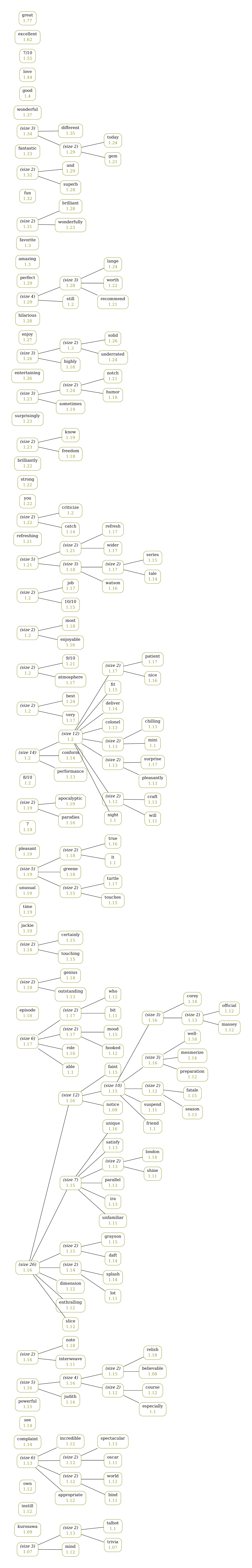}
	\caption{Shows part of the clustering result of the proposed method on IMDB for the class good movie.
		\label{fig:IMDB_proposed_good_movie}}
\end{figure}

\begin{figure}[t]
	\centering
	\includegraphics[trim=0 1720 0 30,clip=true,scale=0.6]{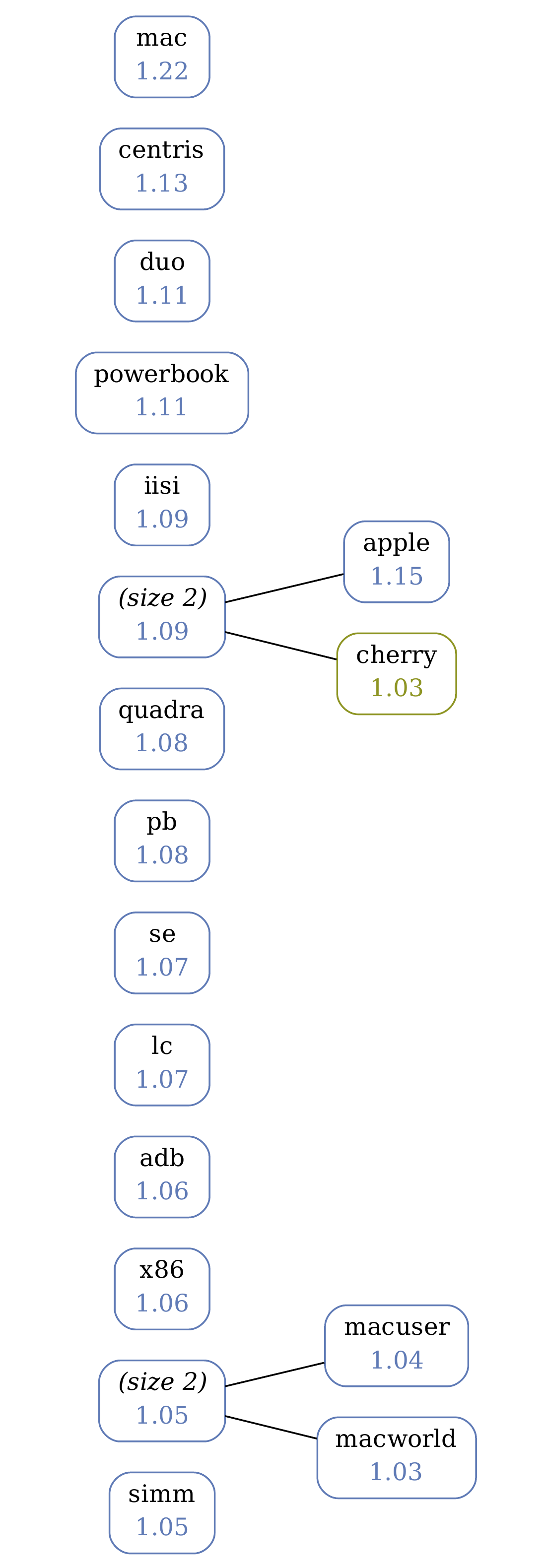}
	\caption{Shows part of the clustering result of the $k$-means clustering on Newsgroup20 for the class Macintosh (hardware).
		\label{fig:newsgroup20all_kMeansClustering_comp_sys_mac_hardware}}
\end{figure}

\begin{figure}[t]
	\centering
	\includegraphics[trim=0 2650 250 30,clip=true,scale=0.5]{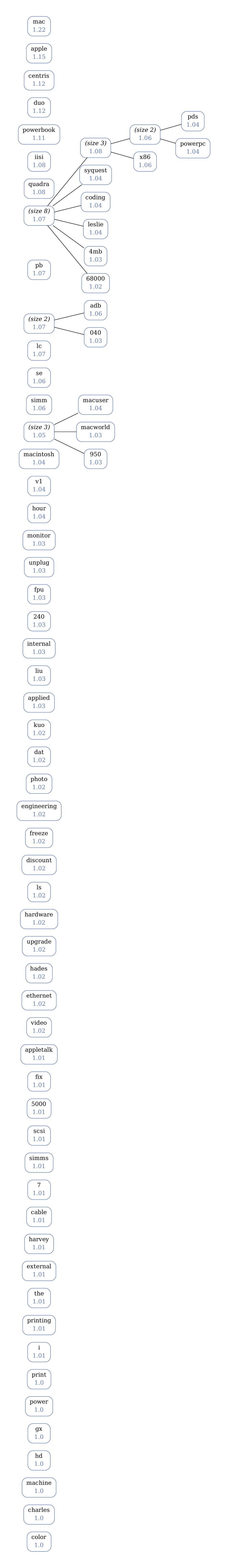}
	\caption{Shows part of the clustering result of the proposed method on Newsgroup20 for the class Macintosh (hardware).
		\label{fig:newsgroup20all_proposed_comp_sys_mac_hardware}}
\end{figure}

Like convex clustering, our proposed method can be used to create a hierarchical clustering structure by varying the hyper-parameter $\nu$. 
In particular, for large enough $\nu$, all covariates collapse to one cluster. On the other side, for $\nu = 0$, each covariate is assigned to a separate cluster. 
In order to ease interpretability, we limit the visualization of the hierarchical clustering structure to all clusterings with $\nu \leq \nu_0$, where $\nu_0$ corresponds to the clustering that was selected with the marginal likelihood criterion. Furthermore, for the visualization of IMDB, we only show the covariates with odds ratios larger than 1.1.
Part of the clustering results of the proposed method, convex clustering and $k$-means are shown in Figures \ref{fig:IMDB_kMeansClustering_bad_movie}, \ref{fig:IMDB_convexClustering_bad_movie}, 
\ref{fig:IMDB_proposed_bad_movie}, \ref{fig:IMDB_proposed_good_movie}, 
and Figures \ref{fig:newsgroup20all_kMeansClustering_comp_sys_mac_hardware}, \ref{fig:newsgroup20all_proposed_comp_sys_mac_hardware}, for IMDB and Newsgroup20, respectively.\footnote{For comparison, for $k$-means and convex clustering we choose (around) the same number of clusters as the proposed method.}

\begin{figure}[t]
	\centering
	\includegraphics[scale=0.5]{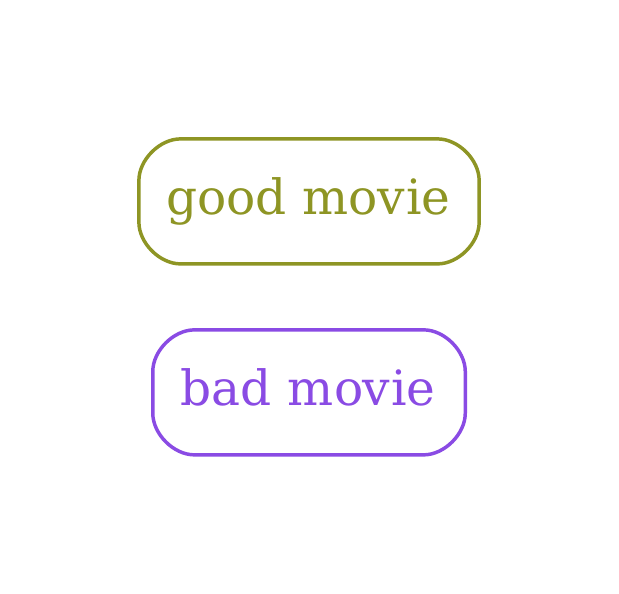}
	\caption{Coloring of each class in IMDB.
		\label{fig:IMDB_legend}}
\end{figure}

\begin{figure}[t]
	\centering
	\includegraphics[scale=0.5]{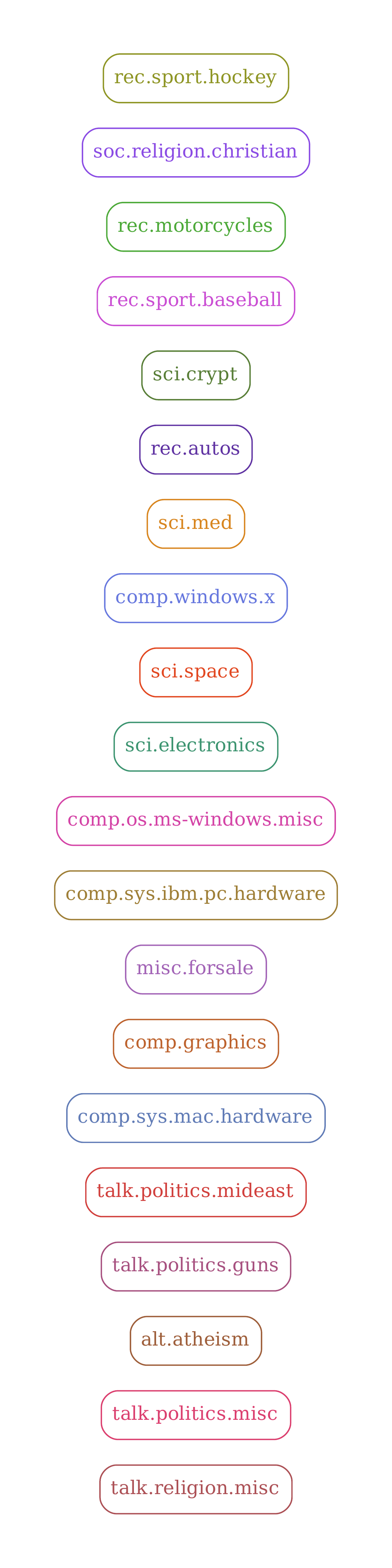}
	\caption{Coloring of each class in Newsgroup20.
		\label{fig:newsgroup20all_legend}}
\end{figure}

Each node represents one cluster. If the cluster is a leaf node, we show the covariate, otherwise we show the size of the cluster (in the following called cluster description). 
From the logistic regression model of each clustering, we associate each cluster to the class with highest entry in weight matrix $B$, and then calculate the odds ratio to the second largest weight. 
The color of a node represents its associated class. All classes with coloring are explained in Figure \ref{fig:IMDB_legend} and \ref{fig:newsgroup20all_legend} for IMDB and Newsgroup20, respectively. 
The odds ratio of each cluster is shown below the cluster description. 


For IMDB, the qualitative differences between the proposed method and $k$-means are quite obvious. 
$k$-means clustering tends to produce clusters that include covariates that are associated to different classes.
An example is shown in Figure \ref{fig:IMDB_kMeansClustering_bad_movie}, where all number ratings are clustered together. 
However, these have very different meaning, since for example 7/10, 8/10, 9/10, 10/10 rate good movies and 
0, 1/10, *1/, 2/10,... rate bad movies. We observe a similar result for convex clustering as shown in Figure \ref{fig:IMDB_convexClustering_bad_movie}.
In contrast, our proposed method distinguishes them correctly by assigning them to different clusters associated with different classes (see Figure 
\ref{fig:IMDB_proposed_bad_movie} and \ref{fig:IMDB_proposed_good_movie}).

For Newsgroup20, the qualitative differences of the proposed method with $k$-means are more subtle, but still we can identify some interesting differences.
For example, $k$-means assigns ``apple'' and ``cherry'' to the same cluster which is 
associated with the class ``Macintosh'' (see Figure \ref{fig:newsgroup20all_kMeansClustering_comp_sys_mac_hardware}). 
In contrast, our proposed method correctly distinguishes these two concepts in the context of ``Macintosh'' (see Figure \ref{fig:newsgroup20all_proposed_comp_sys_mac_hardware}).


\bibliographystyle{plainnat}
\bibliography{all_papers_bibliography_extended}

\begin{thebibliography}{23}
\providecommand{\natexlab}[1]{#1}
\providecommand{\url}[1]{\texttt{#1}}
\expandafter\ifx\csname urlstyle\endcsname\relax
  \providecommand{\doi}[1]{doi: #1}\else
  \providecommand{\doi}{doi: \begingroup \urlstyle{rm}\Url}\fi

\bibitem[Akshay~Agrawal and Boyd(2018)]{cvxpy_rewriting}
Steven~Diamond Akshay~Agrawal, Robin~Verschueren and Stephen Boyd.
\newblock A rewriting system for convex optimization problems.
\newblock \emph{Journal of Control and Decision}, 5\penalty0 (1):\penalty0
  42--60, 2018.

\bibitem[Ando(2010)]{ando2010bayesian}
Tomohiro Ando.
\newblock \emph{Bayesian model selection and statistical modeling}.
\newblock CRC Press, 2010.

\bibitem[Arthur and Vassilvitskii(2007)]{arthur2007k}
David Arthur and Sergei Vassilvitskii.
\newblock k-means++: The advantages of careful seeding.
\newblock In \emph{Proceedings of the Annual ACM-SIAM Symposium on Discrete
  Algorithms}, pages 1027--1035, 2007.

\bibitem[Boyd et~al.(2011)Boyd, Parikh, Chu, Peleato, and
  Eckstein]{boyd2011distributed}
Stephen Boyd, Neal Parikh, Eric Chu, Borja Peleato, and Jonathan Eckstein.
\newblock Distributed optimization and statistical learning via the alternating
  direction method of multipliers.
\newblock \emph{Foundations and Trends{\textregistered} in Machine Learning},
  3\penalty0 (1):\penalty0 1--122, 2011.

\bibitem[Byrd et~al.(1995)Byrd, Lu, Nocedal, and Zhu]{byrd1995limited}
Richard~H Byrd, Peihuang Lu, Jorge Nocedal, and Ciyou Zhu.
\newblock A limited memory algorithm for bound constrained optimization.
\newblock \emph{SIAM Journal on Scientific Computing}, 16\penalty0
  (5):\penalty0 1190--1208, 1995.

\bibitem[Chi and Lange(2015)]{chi2015splitting}
Eric~C Chi and Kenneth Lange.
\newblock Splitting methods for convex clustering.
\newblock \emph{Journal of Computational and Graphical Statistics}, 24\penalty0
  (4):\penalty0 994--1013, 2015.

\bibitem[Collobert et~al.(2011)Collobert, Weston, Bottou, Karlen, Kavukcuoglu,
  and Kuksa]{collobert2011natural}
Ronan Collobert, Jason Weston, L{\'e}on Bottou, Michael Karlen, Koray
  Kavukcuoglu, and Pavel Kuksa.
\newblock Natural language processing (almost) from scratch.
\newblock \emph{The Journal of Machine Learning Research}, 12:\penalty0
  2493--2537, 2011.

\bibitem[Diamond and Boyd(2016)]{cvxpy}
Steven Diamond and Stephen Boyd.
\newblock {CVXPY}: A {P}ython-embedded modeling language for convex
  optimization.
\newblock \emph{Journal of Machine Learning Research}, 17\penalty0
  (83):\penalty0 1--5, 2016.

\bibitem[Domahidi et~al.(2013)Domahidi, Chu, and Boyd]{bib:Domahidi2013ecos}
A.~Domahidi, E.~Chu, and S.~Boyd.
\newblock {ECOS}: {A}n {SOCP} solver for embedded systems.
\newblock In \emph{European Control Conference (ECC)}, pages 3071--3076, 2013.

\bibitem[Hallac et~al.(2015)Hallac, Leskovec, and Boyd]{hallac2015network}
David Hallac, Jure Leskovec, and Stephen Boyd.
\newblock Network lasso: Clustering and optimization in large graphs.
\newblock In \emph{Proceedings of the 21th ACM SIGKDD International Conference
  on Knowledge Discovery and Data Mining}, pages 387--396. ACM, 2015.

\bibitem[Hastie et~al.(2009)Hastie, Tibshirani, and
  Friedman]{hastie2009elements}
Trevor Hastie, Robert Tibshirani, and Jerome Friedman.
\newblock \emph{The elements of statistical learning: data mining, inference,
  and prediction}.
\newblock Springer Science \& Business Media, 2009.

\bibitem[Hastie et~al.(2015)Hastie, Tibshirani, and
  Wainwright]{hastie2015statistical}
Trevor Hastie, Robert Tibshirani, and Martin Wainwright.
\newblock \emph{Statistical learning with sparsity: the lasso and
  generalizations}.
\newblock CRC press, 2015.

\bibitem[Hocking et~al.(2011)Hocking, Joulin, Bach, and
  Vert]{hocking2011clusterpath}
Toby~Dylan Hocking, Armand Joulin, Francis Bach, and Jean-Philippe Vert.
\newblock Clusterpath an algorithm for clustering using convex fusion
  penalties.
\newblock In \emph{28th international conference on machine learning}, page~1,
  2011.

\bibitem[Ledoit and Wolf(2004)]{ledoit2004well}
Olivier Ledoit and Michael Wolf.
\newblock A well-conditioned estimator for large-dimensional covariance
  matrices.
\newblock \emph{Journal of Multivariate Analysis}, 88:\penalty0 365--411, 2004.

\bibitem[Maas et~al.(2011)Maas, Daly, Pham, Huang, Ng, and
  Potts]{maas2011learning}
Andrew~L Maas, Raymond~E Daly, Peter~T Pham, Dan Huang, Andrew~Y Ng, and
  Christopher Potts.
\newblock Learning word vectors for sentiment analysis.
\newblock In \emph{Proceedings of the Annual Meeting of the Association for
  Computational Linguistics}, pages 142--150, 2011.

\bibitem[Meinshausen(2007)]{meinshausen2007relaxed}
Nicolai Meinshausen.
\newblock Relaxed lasso.
\newblock \emph{Computational Statistics \& Data Analysis}, 52\penalty0
  (1):\penalty0 374--393, 2007.

\bibitem[Mikolov et~al.(2013)Mikolov, Sutskever, Chen, Corrado, and
  Dean]{mikolov2013distributed}
Tomas Mikolov, Ilya Sutskever, Kai Chen, Greg~S Corrado, and Jeff Dean.
\newblock Distributed representations of words and phrases and their
  compositionality.
\newblock In \emph{Advances in Neural Information Processing Systems}, pages
  3111--3119, 2013.

\bibitem[O’Donoghue et~al.(2016)O’Donoghue, Chu, Parikh, and
  Boyd]{o2016conic}
Brendan O’Donoghue, Eric Chu, Neal Parikh, and Stephen Boyd.
\newblock Conic optimization via operator splitting and homogeneous self-dual
  embedding.
\newblock \emph{Journal of Optimization Theory and Applications}, 169\penalty0
  (3):\penalty0 1042--1068, 2016.

\bibitem[Pennington et~al.(2014)Pennington, Socher, and
  Manning]{pennington2014glove}
Jeffrey Pennington, Richard Socher, and Christopher~D Manning.
\newblock Glove: Global vectors for word representation.
\newblock In \emph{Conference on Empirical Methods on Natural Language
  Processing}, pages 1532--43, 2014.

\bibitem[Schwarz(1978)]{schwarz1978estimating}
Gideon Schwarz.
\newblock Estimating the dimension of a model.
\newblock \emph{The annals of statistics}, 6\penalty0 (2):\penalty0 461--464,
  1978.

\bibitem[She(2010)]{she2010sparse}
Yiyuan She.
\newblock Sparse regression with exact clustering.
\newblock \emph{Electronic Journal of Statistics}, 4:\penalty0 1055--1096,
  2010.

\bibitem[Vincent and Hansen(2014)]{vincent2014sparse}
Martin Vincent and Niels~Richard Hansen.
\newblock Sparse group lasso and high dimensional multinomial classification.
\newblock \emph{Computational Statistics \& Data Analysis}, 71:\penalty0
  771--786, 2014.

\bibitem[Vinh et~al.(2010)Vinh, Epps, and Bailey]{vinh2010information}
Nguyen~Xuan Vinh, Julien Epps, and James Bailey.
\newblock Information theoretic measures for clusterings comparison: Variants,
  properties, normalization and correction for chance.
\newblock \emph{Journal of Machine Learning Research}, 11\penalty0
  (Oct):\penalty0 2837--2854, 2010.

\end{thebibliography}

\end{document}